\documentclass[reprint]{JASAnew}

\usepackage{currvita}
\usepackage{multirow}

\usepackage{xcolor}

\begin{document}

\title[JASA/Performance of a Deep Neural Network at Detecting North Atlantic Right Whale Upcalls]{Performance of a Deep Neural Network at Detecting North Atlantic Right Whale Upcalls}

\author{Oliver S.\ Kirsebom}
\email{Corresponding author: oliver.kirsebom@dal.ca}
\affiliation{Institute for Big Data Analytics, Dalhousie University, Halifax, Nova Scotia B3H 4R2, Canada}

\author{Fabio Frazao}
\affiliation{Institute for Big Data Analytics, Dalhousie University, Halifax, Nova Scotia B3H 4R2, Canada}

\author{Yvan Simard}
\affiliation{Fisheries and Oceans Canada Chair in underwater acoustics applied to ecosystem and marine mammals, Marine Sciences Institute, University of Qu{\'e}bec at Rimouski, Rimouski, Qu{\'e}bec, Canada}
\affiliation{Maurice Lamontagne Institute, Fisheries and Oceans Canada, Mont-Joli, Qu{\'e}bec, Canada}

\author{Nathalie Roy}
\affiliation{Maurice Lamontagne Institute, Fisheries and Oceans Canada, Mont-Joli, Qu{\'e}bec, Canada}

\author{Stan Matwin}
\affiliation{Institute for Big Data Analytics, Dalhousie University, Halifax, Nova Scotia B3H 4R2, Canada}
\affiliation{Institute of Computer Sciences, Polish Academy of Sciences, Warsaw, Poland}

\author{Samuel Giard}
\affiliation{Maurice Lamontagne Institute, Fisheries and Oceans Canada, Mont-Joli, Qu{\'e}bec, Canada}

\preprint{preprint, JASA}

\date{\today}

\noindent This paper is part of a special issue on The Effects of Noise on Aquatic Life

\begin{abstract} 

Passive acoustics provides a powerful tool for monitoring the endangered North Atlantic right whale ({\it Eubalaena glacialis}), but robust detection algorithms are needed to handle diverse and variable acoustic conditions and differences in recording techniques and equipment. Here, we investigate the potential of deep neural networks for addressing this need. ResNet, an architecture commonly used for image recognition, is trained to recognize the time-frequency representation of the characteristic North Atlantic right whale upcall. The network is trained on several thousand examples recorded at various locations in the Gulf of St.\ Lawrence in 2018 and 2019, using different equipment and deployment techniques. Used as a detection algorithm on fifty 30-minute recordings from the years 2015--2017 containing over one thousand upcalls, the network achieves recalls up to 80\%, while maintaining a precision of 90\%. Importantly, the performance of the network improves as more variance is introduced into the training dataset, whereas the opposite trend is observed using a conventional linear discriminant analysis approach. Our work demonstrates that deep neural networks can be trained to identify North Atlantic right whale upcalls under diverse and variable conditions with a performance that compares favorably to that of existing algorithms.

\end{abstract}

\maketitle

\section{\label{sec:1} Introduction}

The North Atlantic right whale (NARW, {\it Eubalaena glacialis}) comprises a small cetacean population that counted $\sim 400$ individuals in 2017 \citep{pace2017, hayes2018, pettis2019}. Listed as endangered in Canada \citep{cosewic2013}, this population has been declining since 2010 \citep{pettis2019}. NARW used to be mainly distributed along the US continental shelf, up to the Bay of Fundy and the Western Scotian shelf in Canada. This distribution, however, has changed in the last decade \citep{davis2017}. The occasional yearly occurrence of a few individuals in the Gulf of St.\ Lawrence in summer and fall markedly increased in 2015 and a high seasonal occurrence has continued since \citep{simard2019}. This area is a site of intensive fixed-gear fishing. It is crossed by the main continental seaway that connects the Atlantic and the Great Lakes \citep{simard2014}. In 2017, 12 individuals died in the Gulf of St.\ Lawrence. The mortalities involved collisions with ships and entanglement in fixed fishing gear. Protection measures, which include vessel speed reduction and fishing closure, were then put in place by the management authorities in an effort to prevent the recurrence of such events \citep{dfo2018}. 

The key information required to trigger vessel speed reduction and fishing closure is the presence of the animals in the highest risk areas. This information can be acquired over large areas for short time windows from systematic or opportunistic sightings from aircrafts or vessels. However, to obtain continuous round-the-clock information over the season, NARW detection with passive acoustic monitoring (PAM) systems is needed \citep{simard2019}. Various configurations of PAM systems are possible for small- to large-scale coverages \citep{gervaise2019b}, including some supporting detection in quasi real-time such as Viking-WOW buoys (\href{https://ogsl.ca/viking/}{https://ogsl.ca/viking/}), Slocum gliders and fixed buoys \citep{baumgartner2013, baumgartner2019}. 

NARW upcall detection and classification (DC) algorithms are a key component of such PAM systems. Several algorithms, exploiting the time-frequency structure of the call, have been used so far \citep{gillespie2004, mellinger2004, urazghildiiev2006, urazghildiiev2007, urazghildiiev2008, baumgartner2011, simard2019}. Their performance is dependent of the signal to noise ratio (SNR), which varies with the range of the calling whale, the noise levels and the other biological and instrumentation factors \citep{gervaise2019b, simard2019}. The DC performance of these classical signal processing methods under actual {\it in situ} recording conditions tends to plateau around a detection probability of about 50\% (i.e.\ recall index) when the false detection probability is kept below about 10\% \citep{simard2019, dclde2013}. The objective of the present study is to test if modern machine-learning approaches can break this apparent DC performance ceiling. 

Within the last decade, artificial neural networks have become the preferred machine-learning approach for solving a wide range of tasks, outperforming existing computational methods and achieving human-level accuracy in domains such as image analysis~\citep{he2015} and natural speech processing \citep{hinton2012}. %
Originally inspired by the human brain, neural networks consist of a large number of interconnected ``neurons'', each typically performing a simple linear operation on input data, specified by a set of weights and a bias, followed by an activation function. In a supervised training approach, the network is given examples of labeled data, and the weights and biases are adjusted to produce the desired output using an optimization algorithm. 
Modern neural networks exhibit multi-layer architectures, which enable them to build complex concepts out of simpler concepts and hence learn a non-linear representation of the data conducive to solving a given task. Therefore, modern neural networks are often referred to as deep neural networks (DNNs), and the strategy of representing complex data as a nested hierarchy of concepts is referred to as deep learning. %
Two of the most commonly encountered basic architectures are convolutional neural networks (CNNs) and recurrent neural networks (RNNs), which are particularly well adapted to the tasks of analyzing image data and sequential data, respectively. %
The availability of large labeled datasets, containing millions of labeled examples, has been a key factor in the success of DNNs in domains such as image analysis and natural speech processing. Therefore, much of the current research in deep learning focuses on how to train DNNs more efficiently on smaller datasets. %

Shallow neural networks have been employed for the purpose of sound classification in marine bioacoustics since the 1990s, usually combined with a method of feature extraction, e.g.\ \citep{bahoura2010}, but also acting directly on the spectrogram \citep{potter1994, halkias2013}. In the last few years, the first studies employing modern DNNs have been reported. Examples include classification of fish sounds \citep{malfante2018}, detection and classification of orca vocalizations \citep{bergler2019}, classification of multiple whale species \citep{mcquay2017, thomas2019}, and detection and classification of sperm whale vocalizations \citep{bermant2019}. In all cases, CNNs have been leveraged to analyze the information encoded in spectrograms, which is also the strategy adopted in the present work. 

The paper is structured as follows. In Sec.~\ref{sec:2} we first describe how the acoustic data was collected, then discuss the generation of training datasets, the neural network design and the training protocol. In Sec.~\ref{sec:3} we present the results of the detection and classification tasks, which are then discussed in Sec.~\ref{sec:4}. Finally, in Sec.~\ref{sec:5} we summarize and conclude.

\section{\label{sec:2} Materials and Methods}
\subsection{\label{subsec:2:1} Acoustic Data}

\begin{figure*}[ht]
\includegraphics[width=\linewidth]{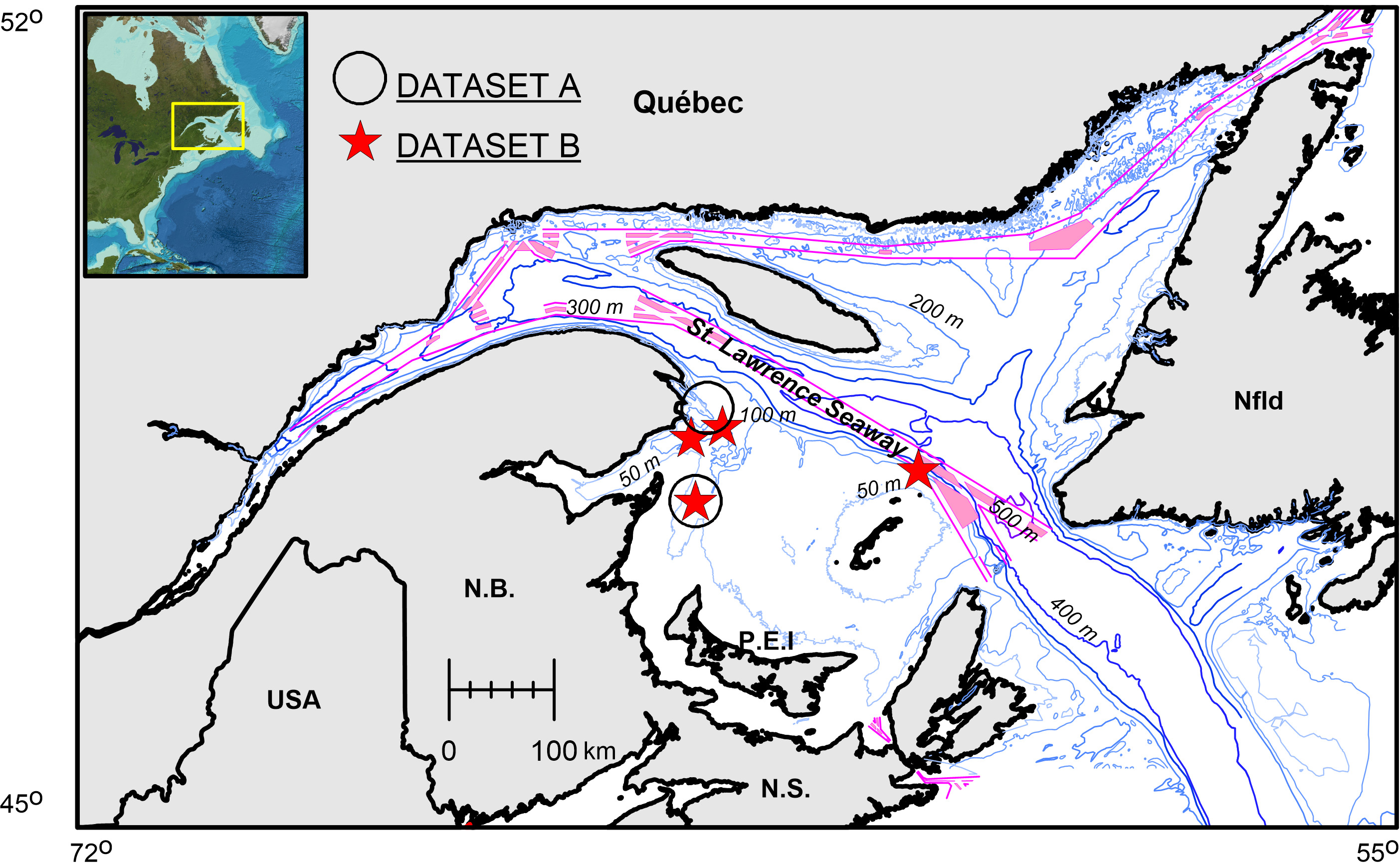}
\caption{\label{fig:map} (Color online) Map of the Gulf of St.\ Lawrence with bathymetry and seaways, showing the location of the PAM stations. Circles indicate surface deployments (dataset A) while stars indicate bottom deployments (datasets B and B$^{\ast}$).}
\end{figure*}

The PAM data were collected between 2015 and 2019 at 6 stations in the southern Gulf of St.\ Lawrence (Fig.~\ref{fig:map}). %
Two different deployment configurations were employed, producing distinct datasets, A, B, and B$^{\ast}$ (Table~\ref{tab:datasets-summary}). %
In addition to these datasets, we have also considered a third dataset, C, which is a subset of the DCLDE 2013 dataset generated from PAM data collected in the Gulf of Maine in 2009.\footnote{\href{https://soi.st-andrews.ac.uk/static/soi/dclde2013/documents/WorkshopDataset2013.pdf}{https://soi.st-andrews.ac.uk/static/soi/dclde2013/documents/\\WorkshopDataset2013.pdf}}

In the case of dataset A, the PAM system was deployed from the surface, with the hydrophone tethered to a real-time ocean observing Viking buoy (Multi-Electronique Inc., Rimouski, Qc, Canada, http://www.multi-electronique.com/buoy.html) with a 60-m long cable floating at the surface for half of its length. The recording digital hydrophone, at a depth of $\sim 30$~m, was an ic-Listen HF (Ocean Sonics, Truro Heights, N.S., Canada, \href{https://oceansonics.com/product-types/iclisten-smart-hydrophones/}{https://oceansonics.com/product-types/iclisten-smart-hydrophones/}). It sampled continuously the raw (0 gain) acoustic signal with 24-bit resolution. The receiving sensitivity of the hydrophone was $-170$~dB~re~1~V~$\mu$Pa$^{-1}$. 

In the case of datasets B and B$^{\ast}$, the PAM system used short ($< 10$~m) ``I''-type oceanographic moorings, with an anchor, an acoustic release, and streamlined underwater floats, for bottom deployment at depths varying from 75~m to 125~m, with the autonomous hydrophone $\sim 5$~m above the seafloor. The recording equipment consisted of AURAL M2 (Multi-Electronique Inc., Rimouski, Qc, Canada, http://www.multi-electronique.com/aural.html) sampling the 16-dB pre-amplified acoustic signal with a 16-bit resolution at duty cycles of 15~min or 30~min every hour. The receiving sensitivity of the HTI 96-MIN (High Tech Inc., Gulfport, MS) hydrophone equipping the AURAL is $-164\pm 1$~dB~re~1~V~$\mu$Pa$^{-1}$ over the $< 0.5$-kHz bandwidth used here. Further details can be found in \citep{simard2019}. 

Because of the different acoustic equipment and deployment types, the recordings from the two datasets differed significantly in terms of signal amplitude and noise background from different sources, including flow noise, strum, and knocks resulting from the effects of tidal currents and the surface motion due to waves on the hydrophone and deployment apparatus. Additional SNR variability of the recordings is introduced by the different locations and depths at which the hydrophones were deployed in the southern Gulf of St.\ Lawrence, providing different exposures to the above environmental conditions and to the shipping noise field from the main seaway \citep{aulanier2016, simard2019}. The datasets used here therefore represent a large range of conditions that can be encountered in realizing the DC task for the low-frequency NARW upcall from acoustic data collected using different PAM systems. To develop a deep learning model that is robust to such realistic range of variability, no effort was made to enhance the SNR before feeding the data to the neural network. 

\begin{table*}[ht]
\caption{\label{tab:datasets-summary}Datasets used in this work.}
\begin{ruledtabular}
\begin{tabular}{ccccc}
Dataset  &  Deployment type  &  Location  &  Year(s)  &  Analysis method\\
\hline
  A  &  Surface buoy    &  Gulf of St.\ Lawrence  &  2019        &  Expert validation of detections reported by TFBD algorithm\\
  B  &  Bottom mooring  &  Gulf of St.\ Lawrence  &  2018        &  Expert validation of detections reported by TFBD algorithm\\
  B$^{\ast}$  &  Bottom mooring  &  Gulf of St.\ Lawrence  &  2015--2017  &  Full manual analysis of fifty 30-min recordings \\
  C  &  Bottom mooring  &  Gulf of Maine          &  2009        &  Full manual analysis of seven days of recordings \\
\end{tabular}
\end{ruledtabular}
\end{table*}

\subsection{\label{subsec:2:2} Training and Test Datasets}

Datasets A and B were first analyzed with a classical time-frequency based detector (TFBD) following \citep{mellinger2004} and \citep{mouy2009}. This algorithm looks for a typical image of NARW upcall in the SNR-enhanced  (i.e., noise-subtracted), high-resolution ($32\textrm{ ms}\times 3.9$~Hz) spectrogram of the recordings, and a detection is triggered by the degree of cross-coincidence. The NARW upcall template used was a 1-s, 100--200~Hz chirp with a $\pm 10$~Hz bandwidth. For further details, see \citep{simard2019}. The resulting detections were then manually validated by an expert by examining the spectrogram and labelled as ``true'' or ``false'' using the longer call pattern context in a $\sim 1$-min window. In NARW call occurrence studies, the false detections are then eliminated. For the present work, however, both true and false detections were extracted from the recordings and used as ``positives'' and ``negatives'', respectively, for building the training datasets A and B (Table~\ref{tab:datasets}). %
For dataset C, we have used the existing annotations from the DCLDE 2013 Challenge. Finally, we have built the composite datasets AB and ABC by combining all the samples from the individual datasets.

\begin{table*}[ht]
\caption{\label{tab:datasets}Number of samples in the datasets used for training and testing the classifiers. The composite datasets AB and ABC were produced by combining all the samples from the individual datasets.}
\begin{ruledtabular}
\begin{tabular}{ccccccc}
 \multirow{2}{*}{Dataset}  &  \multicolumn{3}{c}{Training and validation}  &  \multicolumn{3}{c}{Testing}\\\cline{2-4}\cline{5-7} 
 &  No.\ samples  &  Positives  &  Negatives  &  No.\ samples  &  Positives  &  Negatives \\ 
\hline
  A    &  $1,767$  &  42\%  &  58\%  &  307  &  18\%  &  82\% \\
  B    &  $3,309$  &  61\%  &  39\%  &  579  &  59\%  &  41\% \\
  C    &  $3,000$  &  50\%  &  50\%  &  $-$  &  $-$   &  $-$  \\
  AB   &  $5,076$  &  55\%  &  45\%  &  886  &  45\%  &  55\% \\
  ABC  &  $8,076$  &  53\%  &  47\%  &  $-$  &  $-$   &  $-$  \\
\end{tabular}
\end{ruledtabular}
\end{table*}

To examine the accuracy of the validation protocol used for producing datasets A and B, a second expert was subsequently tasked with reviewing a subset of the annotations. %
The review differed from the validation in several ways: the second expert had knowledge of both the labels assigned by the first expert and the classification proposed by the DNN. The second expert used raw spectrograms while the first expert used SNR-enhanced spectrograms\footnote{SNR enhancement can have two effects: It may allow extracting signals deeply embedded in noise, which cannot be seen in the raw spectrograms, but it may also generate artifacts that are mimicking real signals.} and a larger temporal context, including considerations of occurrence probability over the seasons. Finally, the first expert was instructed to adopt a more conservative annotation strategy, always assigning a negative label in cases of substantial doubt, whereas no such instruction was given to the second expert. %
The results of the annotation review will be discussed in Sec.~\ref{sec:4}.

The extracted segments are 3~s long and centered on the midpoint of the upcall, as determined by the TFBD algorithm. However, the midpoint determined by the algorithm rarely coincides with the actual midpoint of the upcall, producing segments that are misaligned by up to 0.5~s in either direction. We note that such quasi-random time shifts are desirable for training a DNN classifier because they encourage the network to learn a more general, time translation invariant, representation of the upcall. %

\begin{figure}[ht]
\includegraphics[width=\linewidth]{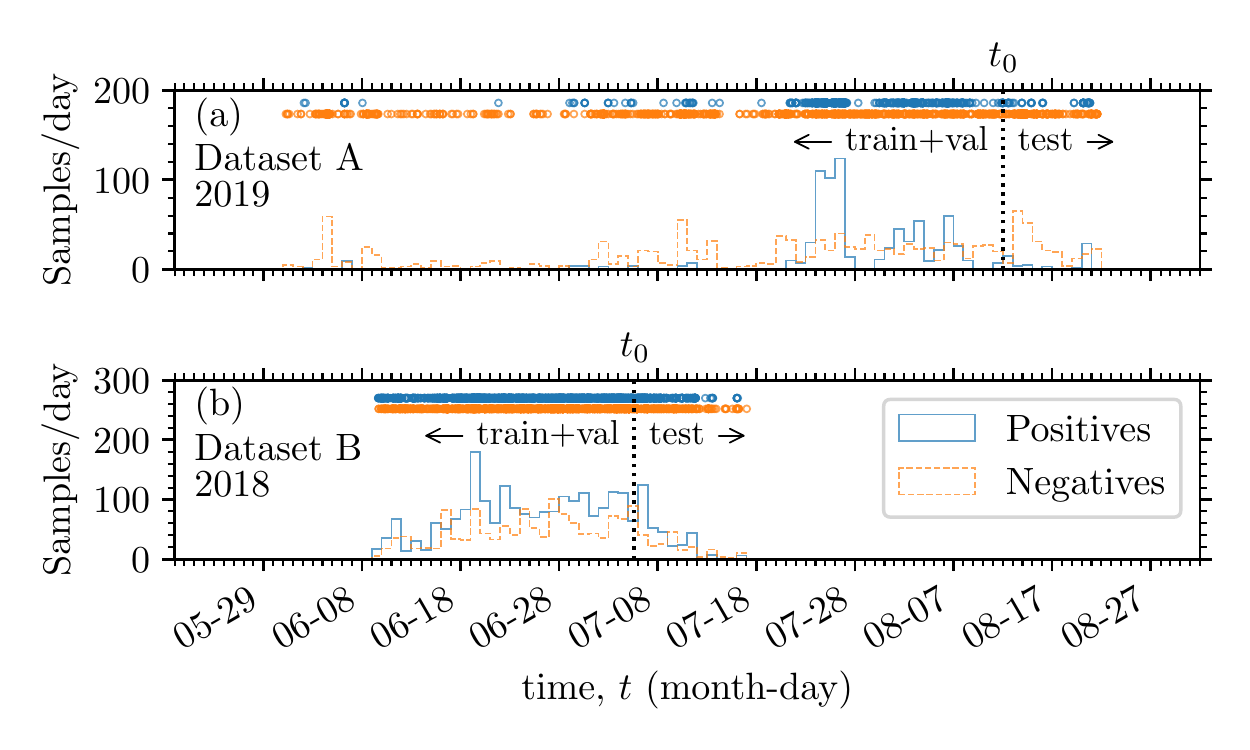}
\caption{\label{fig:timeline} (Color online) Time-split used to produce training and validation sets and test sets for Datasets A (a) and B (b).}
\end{figure}

For the purpose of testing the classification performance of the trained models, including their capacity for generalizing, we split datasets A and B as follows: Samples obtained at times $t < t_0$ were used for training and validation, while samples obtained at times $t > t_0$ were retained for testing. Here, $t_0$ was chosen to produce a 85:15 split ratio between the number of samples used for training and validation and the number of samples used for testing (Fig.~\ref{fig:timeline}). %
This split implies temporal separations of 52~min and 33~min between the latest sample in the training dataset and the earliest sample in the test dataset for A and B, respectively. 

\begin{figure}[ht]
\includegraphics[width=\linewidth]{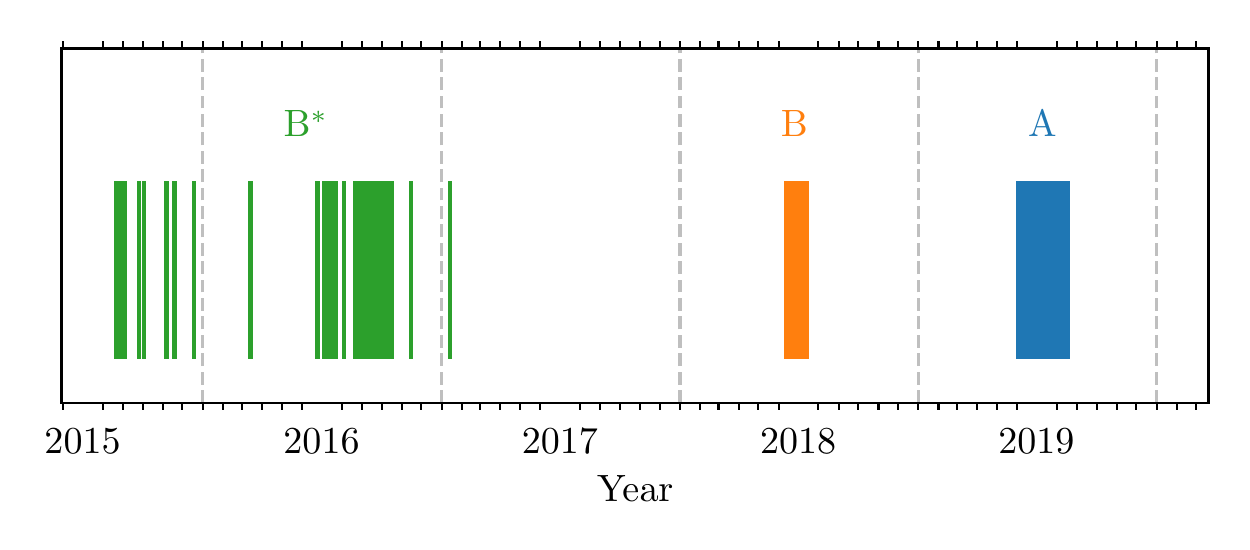}
\caption{\label{fig:timeline-overview} (Color online) Temporal distribution of the training datasets A and B and the continuous test dataset B$^{\ast}$.}
\end{figure}

From the bottom deployments we also produced dataset B$^{\ast}$ consisting of fifty 30-min segments from two years between 2015 and 2017 (Fig.~\ref{fig:timeline-overview}). These data were used for testing the detection performance of the neural network on continuous data. The data cover all seasons and times of the day, hence providing a representative picture of the acoustic conditions found in the Gulf of St.\ Lawrence. Moreover, the data have no temporal overlap with the training datasets A and B, which originate from 2019 and 2018, respectively. Dataset B$^{\ast}$ was manually analyzed in its entire length by a third expert who identified $1,157$ NARW upcall occurrences. %

\subsection{\label{subsec:2:3} Spectrogram and SNR Computation}

First, the 3-s acoustic segments were downsampled to $1,000$ samples~s$^{-1}$ using MATLAB's resample function, which employs a polyphase anti-aliasing filter. %
The spectrogram representation was then computed on a dB scale using a window size of 0.256~s, a step size of 0.032~s (88\% overlap), and a Hamming window. These parameters have been shown to be optimal for identifying NARW upcalls \citep{gervaise2019a} and produce a spectrogram with (time, frequency) dimensions of $94\times 129$. %
We note that the spectrograms were fed to the network in their raw form. In particular, no effort was made to normalize the spectrograms to correct for systematic differences in signal amplitude in the three datasets. This approach was adopted to produce the most general model possible.
%

\begin{figure}[ht]
\includegraphics[width=\linewidth]{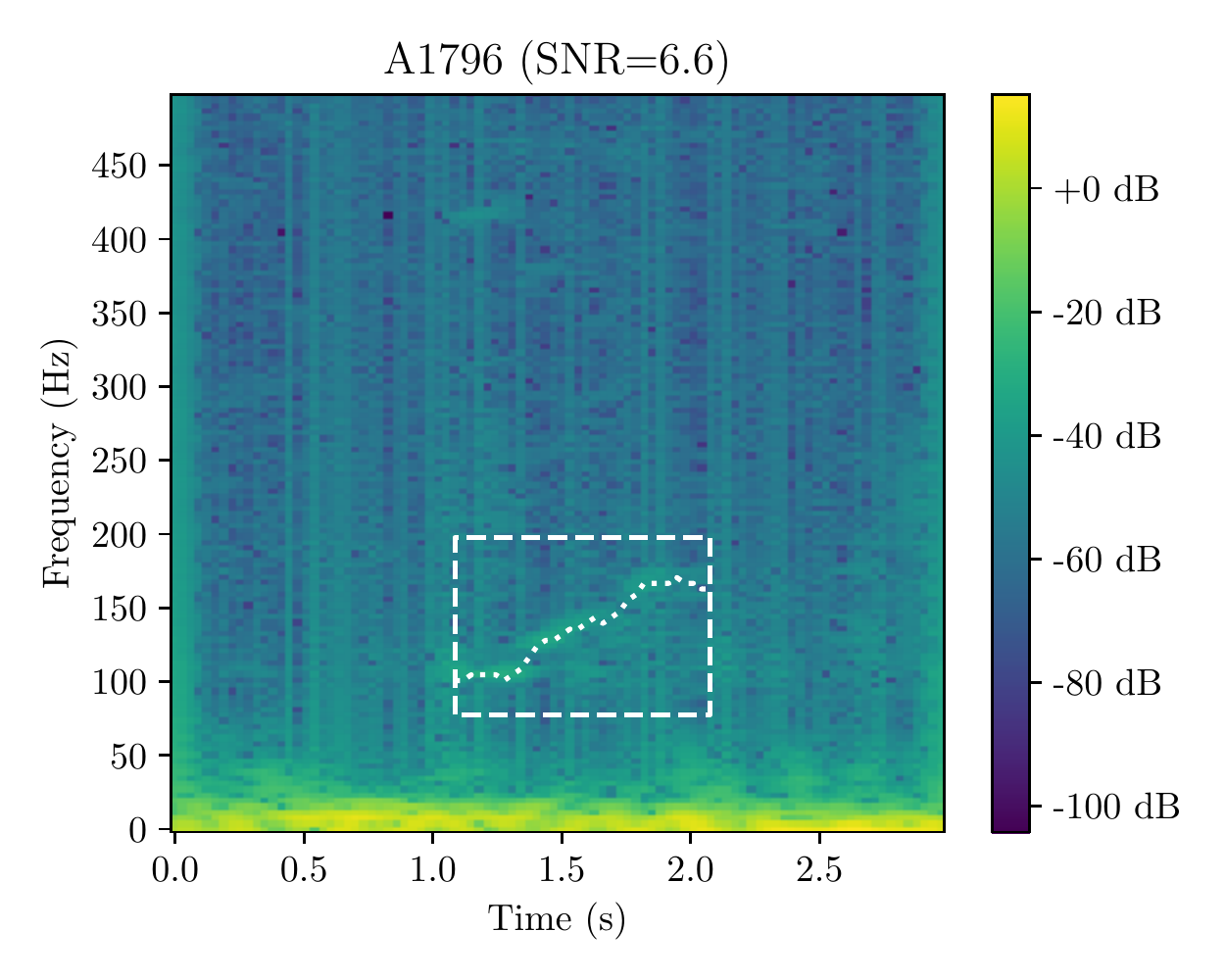}
\caption{\label{fig:snr-algorithm} (Color online) Positive spectrogram sample from dataset A. Superimposed is a 1-s window centered on the upcall (dashed line) and a trace drawn along the upcall (dotted curve), as computed by the heuristic SNR algorithm described in the text.}
\end{figure}

For the estimation of the SNR value of each sample, positive or negative, the following heuristic algorithm was implemented: (1) A denoised spectrogram, $X_d$, was created by subtracting first the median value of each time slice (to reduce broadband, impulsive noise) and then subtracting the median value of each frequency slice (to reduce tonal noise). (2) The mid-point of the upcall was determined by sliding a 1-s wide window across the denoised spectrogram while seeking to maximize,
\begin{equation*} 
\text{sum}_t(\text{max}_f(X_d)) + \text{sum}_f(\text{max}_t(X_d)) \; ,
\end{equation*}
in the frequency interval 80--200~Hz, where the subscripts indicate the axis ($t$:time, $f$:frequency) along which the mathematical operation is applied. (3) A trace was drawn by connecting the pixels with the maximum value in each time slice of $X_d$. (4) The median value of the original spectrogram, $X$, was computed along this trace, including also the three pixels immediately above and below to account for the finite ``width'' of the upcall. (5) Finally, the median values of $X$ in the 0.5-s adjacent windows were computed for the frequency interval 80--200~Hz and subtracted. Fig.~\ref{fig:snr-algorithm} shows the result obtained on a typical positive sample from dataset A. 
We stress that SNR estimation is highly challenging for the datasets considered in this work because of non-uniform stationary noise, transient noise, and distortion of the upcalls due to propagation effects.

\subsection{\label{subsec:2:4} Neural Network Architecture}

The problem was set up as a binary classification task: A neural network was trained to classify the 3-s spectrograms according to the criterion, contains (positive class, 1) or does not contain (negative class, 0) a NARW upcall.  %
We used a residual network (ResNet), which is a CNN architecture mainly built of residual blocks with skip connections \citep{he2016}.
CNNs consist of a stack of convolutional layers followed by a few fully connected layers.
During the training process, the convolutional layers learn to extract patterns from the input images, which are passed to the fully connected layers for classification \citep[Chapter~9]{goodfellow2016}.
The residual blocks in a ResNet are composed of convolutional layers, but allow some connections between layers to be skipped, thereby avoiding ``vanishing'' and ``exploding'' gradients during training \citep{he2016}.
We used blocks with batch normalization \citep{ioffe2015} and rectified linear units (ReLU) \citep{nair2010}.
The architecture was composed of eight such blocks preceded by one convolutional layer and followed by a batch normalization layer, global average pooling \citep{lin2013}, and a fully connected layer with a softmax function, which is responsible for the classification. 
Finally, the output layer gives a score in the range 0--1 for each of the two classes (positive and negative), which add up to 1. %

\subsection{\label{subsec:2:5} Training Protocol}
We trained the network on two NVIDIA RTX 2080 Ti GPUs with 11GB of memory. %
Training was performed with a batch size of 128 and terminated after a pre-set number of epochs, $N$. That is, 128 samples were passed through the network between successive optimizations of the weights and biases, and every sample in the training dataset was passed through the network $N$ times. Weights and biases were optimized with the ADAM optimizer \citep{kingma2014} using the recommended parameters: an initial learning rate of 0.001, decay of 0.01, $\beta_{1}$ of 0.9, and $\beta_{2}$ of 0.999. 
No effort was made to explore the effects of these parameters on the training outcome.
The network was trained to maximize the $F_1$ score, defined as the harmonic mean of precision and recall, $F_1 = 2 P R /(P + R)$,
where $R$ is the recall, i.e., the fraction of the upcalls that were detected, and $P$ is the precision, i.e., the fraction of the detected upcalls that were in fact upcalls. Thus, the $F_1$ score considers both recall and precision and attaches equal importance to the two.

Initially, the network was trained using 5-fold cross-validation with a 85:15 random split between the training and validation sets, allowing us to confirm that the optimization had converged without overfitting. Based on these initial training sessions, $N=100$ was found to provide an optimal choice for all the training datasets. The network was then trained on the full training datasets without cross-validation for $N=100$ epochs. This was repeated nine times with different random number generator seeds to assess the sensitivity of the training outcome to the initial conditions.

\subsection{\label{subsec:2:6} Linear Discriminant Analysis}

To establish a baseline against which to compare the performance of the neural network, we implemented a linear discriminant analysis (LDA) model following the approach of \citep{martinez2001}, noting that such models have traditionally been adopted for solving sound detection and classification tasks in marine bioacoustics. First, the $94\times 129$ spectrogram matrix was flattened to a vector of length $12,126$. Second, the dimensionality was reduced by means of principal component analysis (PCA). Third, we trained the LDA classifier using a least-squares solver combined with automatic shrinkage following the Ledoit-Wolf lemma \citep{Ledoit2004}. The training was repeated for several choices of PCA dimensionality using a 85:15 random split between training and validation sets, and the dimensionality yielding the best performance on the validation set was selected.

\subsection{\label{subsec:2:7} Detection Algorithm}

For the purpose of detecting NARW upcalls in continuous acoustic data, the following simple algorithm was implemented: First, the data were segmented using a window of 3~s and a step size of $\Delta t = 0.5$~s. Each 3-s segment was then fed to the DNN classifier, producing a sequence of classification scores between 0--1, which we interpret as a time-series of upcall occurrence probabilities. Empirically, we found it useful to smoothen the classification scores using a five-bin (2.5~s) wide averaging window. This greatly reduced the number of false positives (factor of $\sim 5$) at the cost of a modest increase in the number of false negatives (factor of $\sim 2$). Finally, we applied a uniform detection threshold, setting the bin value to 1 (``positive'') when the score was greater than or equal to the threshold and 0 (``negative'') when it was below. 

For the computation of recall, precision, and false-positive rate, we merged adjacent positive (1) bins into ``detection events'', which extend from the lower edge of the first bin, $t$, to the upper edge of the last bin, $t + N \Delta t$, $N$ being the number of bins in the event. 
To allow for minor temporal misalignments between annotations and detections, we adopted a temporal buffer of 1.0~s, effectively expanding every detection event to $[t - 2\Delta t, \; t + (N+2) \Delta t]$. %
Considering the primary intended application of the detection algorithm, namely, to quantify upcall occurrences in PAM data and provide occurrence times of sufficient accuracy to aid validation by a human analyst, such a small temporal buffer is fully justifiable. %
The recall was then computed as the fraction of annotated upcalls that exhibit at least 50\% overlap with a detection event, while the precision was computed as the fraction of the detection events that exhibit at least 50\% overlap with an annotated upcall. Any detection event that did not overlap with an annotated upcall or exhibited less than 50\% overlap was counted as one false positive for the computation of the false-positive rate. %

\section{\label{sec:3} Results}

\subsection{\label{subsec:3:1} Classification Performance}

\begin{figure*}[ht]
\includegraphics[width=0.495\linewidth]{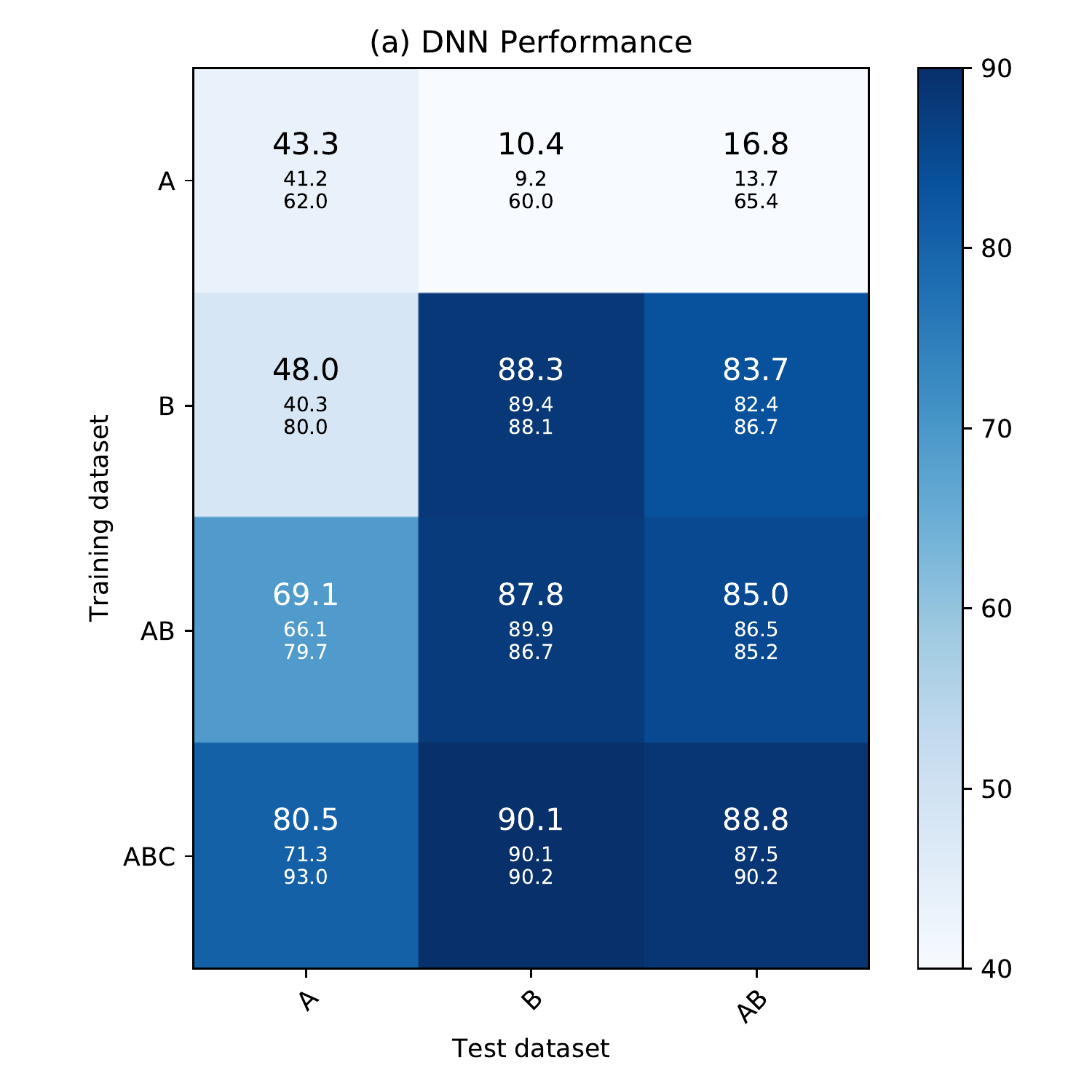}
\includegraphics[width=0.495\linewidth]{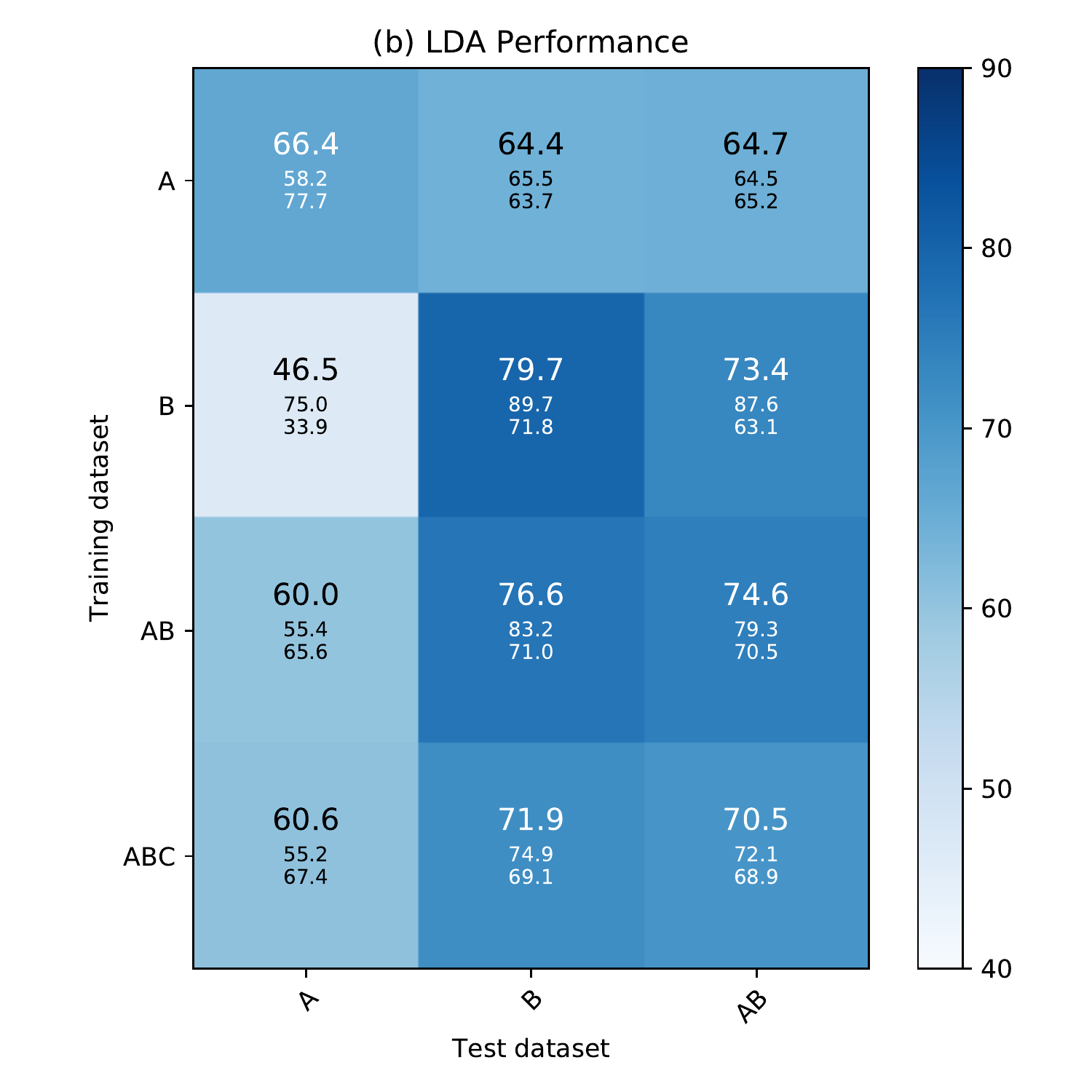}
\caption{\label{fig:DNN-performance} (Color online) (a) DNN classification performance in terms of $F_1$ score (top, large font), recall (middle, small font), and precision (bottom, small font). Rows: training dataset; columns: test dataset. The colorscale indicates the $F_1$ score. (b) LDA classifier performance.}
\end{figure*}

The classification performance of the DNN and LDA classifiers on the test datasets are summarized in terms of the average $F_1$ score, recall, and precision obtained in the nine independent training sessions (Fig.~\ref{fig:DNN-performance}). The DNN model trained on the ABC dataset exhibits the best overall performance, achieving a recall of $87.5\%$ and a precision of $90.2\%$ on the AB test set (with standard deviations of $1.1\%$ and $1.2\%$) and outperforming the baseline LDA model by a statistically significant margin (as evident from Fig.~\ref{fig:DNN-snr-performance} below). 

We have investigated the effect of increasing the size of dataset C by up to a factor of 10 ($15,000$ upcalls), but found only a negligible improvement in the performance of the DNN model. (We note that the DCLDE 2013 dataset contains $6,916$ logged calls. To produce a dataset with $15,000$ upcalls we added time-shifted copies of the logged calls to the dataset.) %

We have also investigated the effect of discarding samples with SNR below a certain minimum value, SNR$_{\mathrm{min}}$, from the AB test set (Fig.~\ref{fig:DNN-snr-performance}). For the DNN model, we observe a gradual increase in performance as we restrict our attention to upcalls with increasingly larger SNR values, with the recall improving from 89\% at $\mathrm{SNR}_{\mathrm{min}}\simeq 0$ to 98\% at $\mathrm{SNR}_{\mathrm{min}}\simeq 12$ and the precision improving from 91\% to 100\% across the same range of SNR. The performance of the LDA model also improves with increasing SNR. This is especially true for the precision, which improves from 70\% at $\mathrm{SNR}_{\mathrm{min}}\simeq 0$ to 95\% at $\mathrm{SNR}_{\mathrm{min}}\simeq 12$, whereas the recall reaches a maximum of 82\% before worsening at the largest SNR values.

\begin{figure}[ht]
\includegraphics[width=\linewidth]{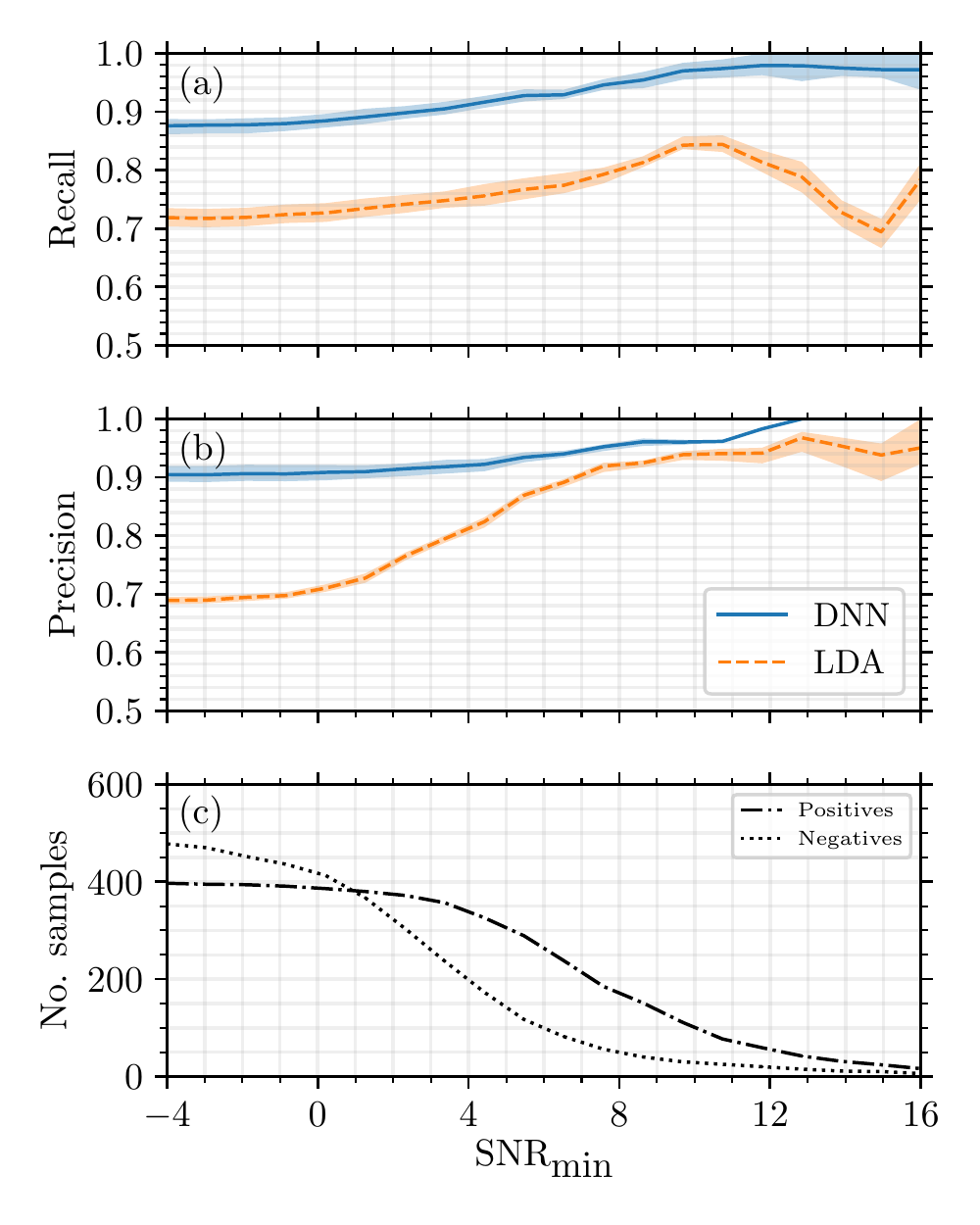}
\caption{\label{fig:DNN-snr-performance} (Color online) (a) Effect of discarding samples with $\mathrm{SNR} < \mathrm{SNR}_{\mathrm{min}}$ from the AB test dataset on the recall of the DNN and LDA models trained on ABC. The lines show the average recall obtained in the nine training sessions, while the shaded bands show the 10\% and 90\% percentiles. (b) Same, for the precision. (c) Number of positive and negative samples in the AB test dataset with $\mathrm{SNR} \geq \mathrm{SNR}_{\mathrm{min}}$.}
\end{figure}

\begin{figure*}[ht]
\includegraphics[width=\linewidth]{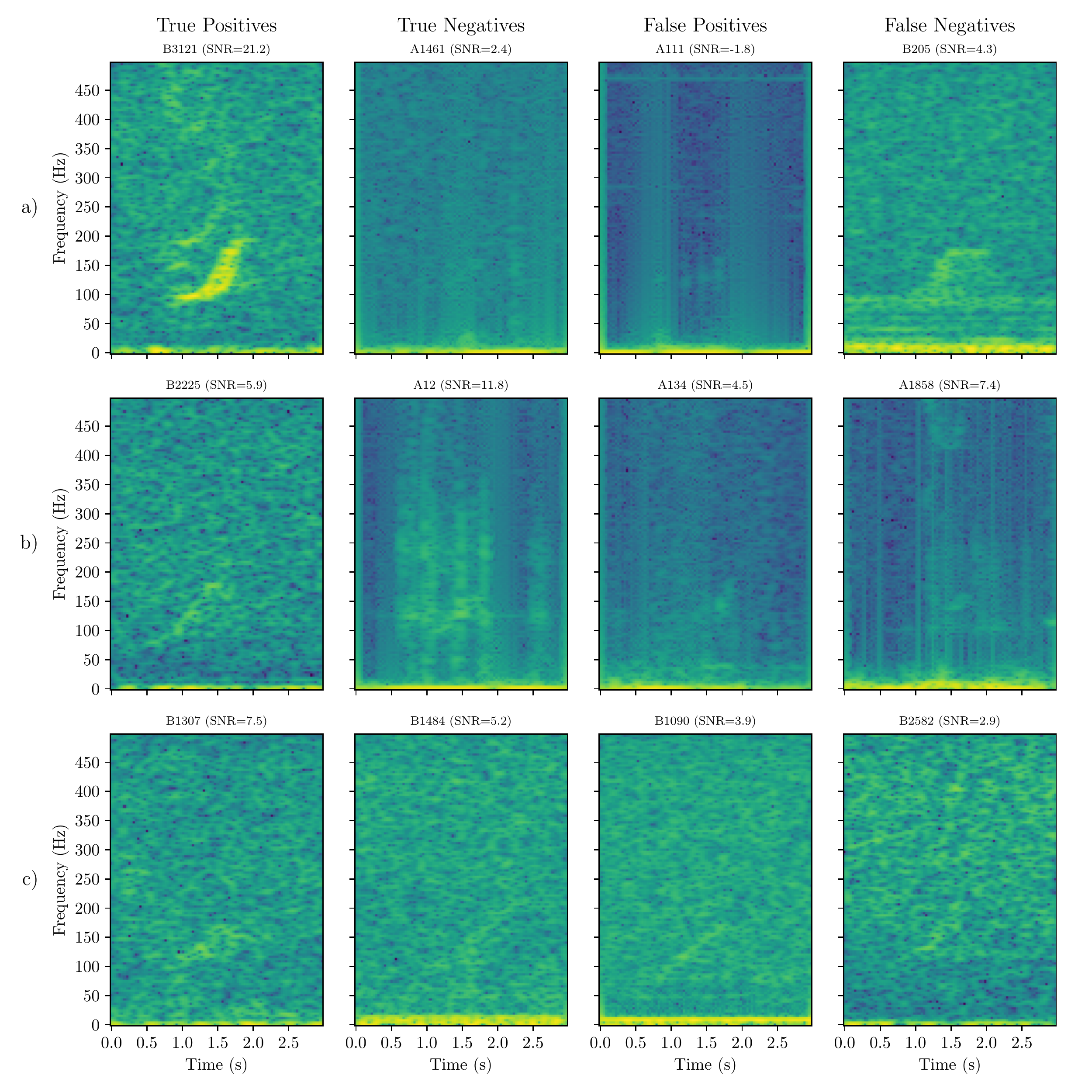}
\caption{\label{fig:spec-examples} (Color online) Representative 3-s spectrogram samples. First column: true positives; second column: true negatives; third column: false positives; fourth column: false negatives. For each category, three examples are given reflecting different levels of certainty and difficulty as perceived by the second expert: (a) certain and easy; (b) certain, but difficult; (c) uncertain. Spectrograms are labeled by their ID and SNR (in dB).}
\end{figure*}

In the following, we examine a small set of representative spectrogram samples, which have been either correctly classified or misclassified by the DNN model (Fig.~\ref{fig:spec-examples}). We divide the samples into true positives, true negatives, false positives, and false negatives, and for each category we give three examples reflecting different levels of certainty and difficulty as perceived by the second expert: (a) certain and easy, (b) certain, but difficult, and (c) uncertain. Here, it must be remembered that the experts had access to a larger temporal context of $\sim 1$~min to inform their decision. Notably, this may have helped the expert to correctly identify calls with low SNR in cases where the calls form part of a call series.

A few observations can be made: the model is able to correctly identify upcalls with very different SNR (B2225, B3121); the model is able to correctly classify negatives containing potentially confusing patterns (A12), but not always (A111); the model struggles in cases with low SNR (A134, A1858); the model can be confused by tonal noises and multipath echoes (B205). %
These deficiencies could potentially be resolved by enlarging the temporal window, thereby giving the model access to the same contextual information that is available to the human analyst, notably the appearance of an upcall series. 

Finally, we highlight a limitation of the classification results reported in this section. Since datasets A and B only contain samples flagged by the TFBD algorithm, the performance demonstrated on these datasets is not necessarily representative of the performance on a random selection of samples from a continuous recording, and the reported metrics (Fig.~\ref{fig:DNN-performance}) cannot be readily applied to continuous data. In the next section, we address this limitation by testing the performance of the classifiers on dataset B$^{\ast}$, which has been subject to full manual analysis.

\subsection{\label{subsec:3:3} Detection Performance on Continuous Data}

The detection algorithm introduced in Sec.~\ref{subsec:2:7} was tested on dataset B$^{\ast}$, which consists of fifty 30-minute segments and has a total of $1,157$ upcalls. The number of calls per file exhibits significant variation, ranging from none to 100 with a median value of 15. %

\begin{figure}[ht]
\includegraphics[width=\linewidth]{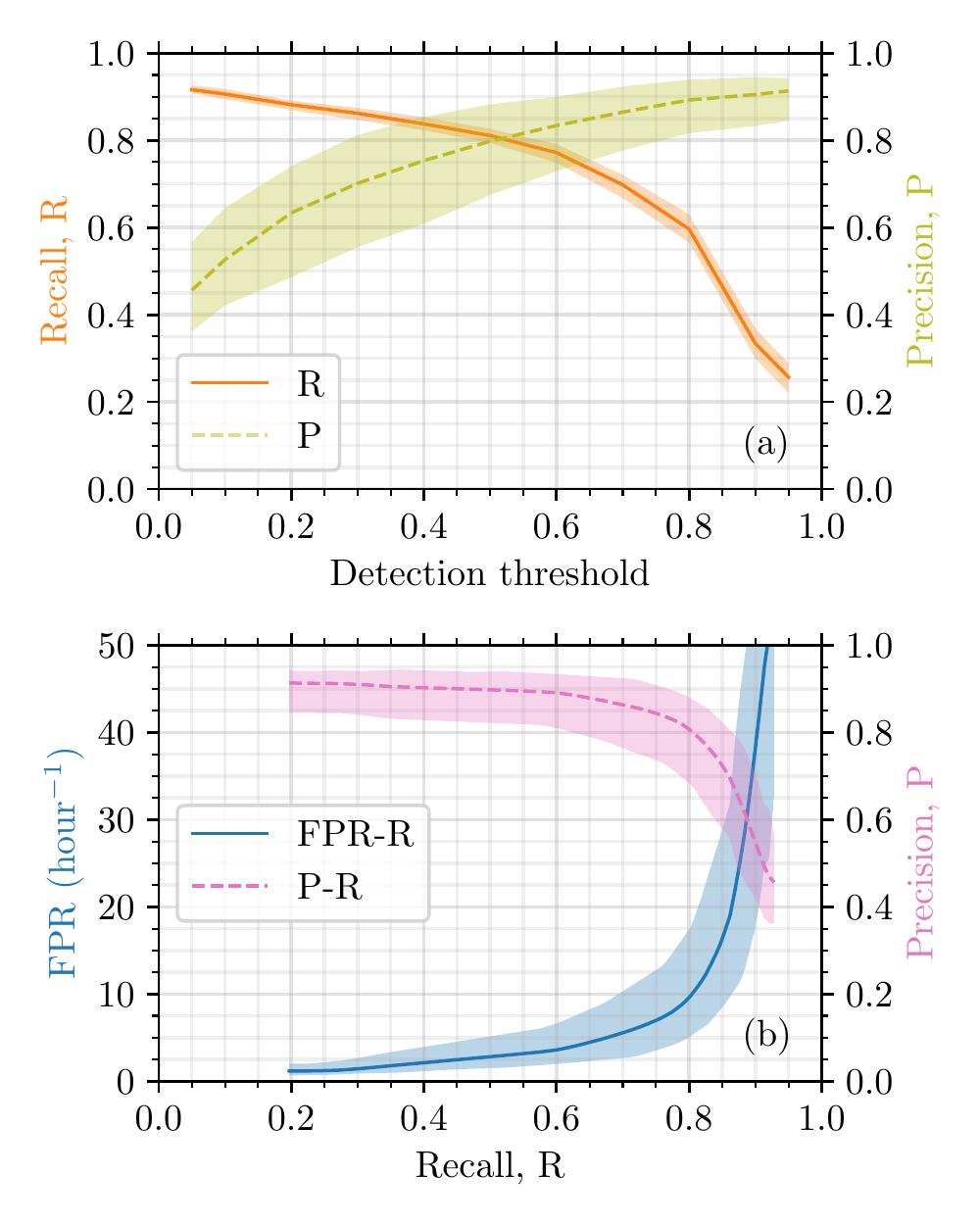}
\caption{\label{fig:detector-performance} (Color online) Detection performance on the continuous test data (dataset B$^{\ast}$) in terms of recall (R), precision (P), and false-positive rate (FPR). The lines show the average performance while the shaded bands show the 10\% and 90\% percentiles. (a) R and P versus the adopted detection threshold. (b) P-R and FPR-R curves.}
\end{figure}

The performance of the detection algorithm is summarized in terms of recall, precision, and false-positive rate (Fig.~\ref{fig:detector-performance}). 
The detection threshold is seen to provide a convenient tunable parameter to adjust the detection performance, depending on whether high precision or high recall is desired. One also notes that the nine independent training sessions produced detectors with very similar recall, but varying levels of precision. In particular, the best-performing detector achieves a recall of 80\%, while maintaining a precision above 90\%, corresponding to a false-positive rate of 5 occurences per hour for this particular test dataset, while the ``average'' detector achieves a recall of 60\% for the same level of precision.

\begin{figure}[ht]
\includegraphics[width=\linewidth]{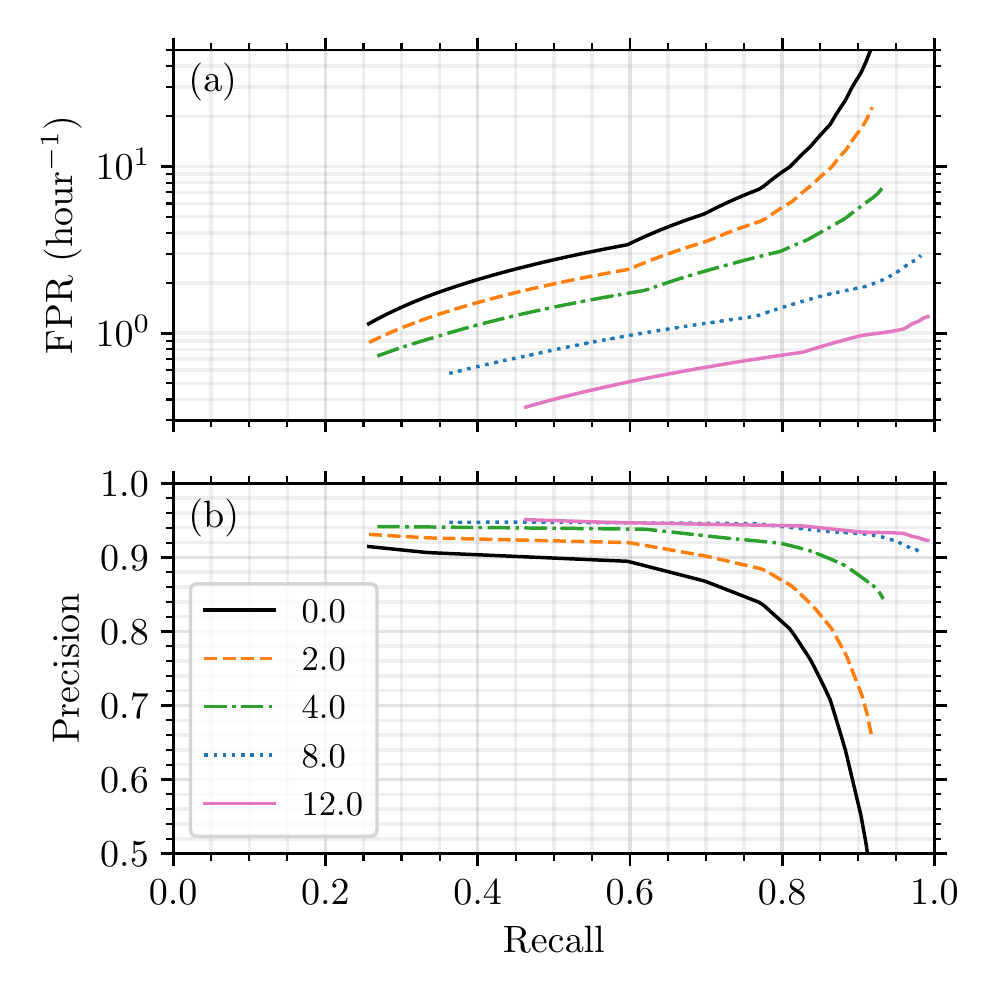}
\includegraphics[width=\linewidth]{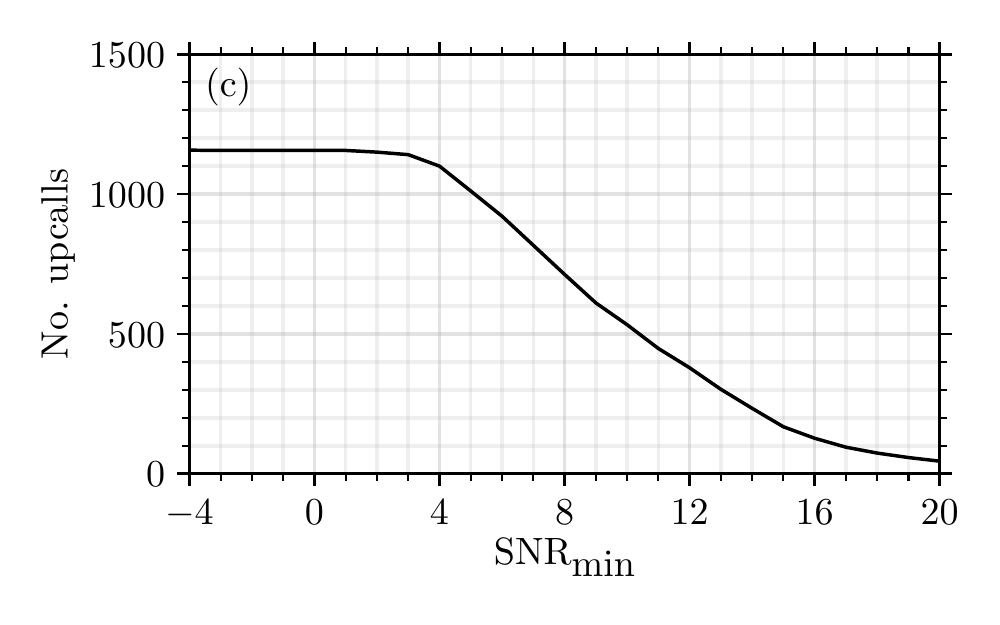}
\caption{\label{fig:detector-snr-performance} (Color online) Detection performance of the ``average'' model on the continuous test data (dataset B$^{\ast}$) for the upcalls meeting the criterion $\mathrm{SNR} > \mathrm{SNR}_{\mathrm{min}}$. The performance is shown in terms of recall (R), precision (P), and false-positive rate (FPR) for the five cut-off values $\text{SNR}_{\text{min}}=0.0$, 2.0, 4.0, 8.0, 12.0. (a) FPR vs.\ R; (b) P vs.\ R; (c) Number of upcalls vs.\ $\text{SNR}_{\text{min}}$.}
\end{figure}

Finally, we have considered the effect of discarding samples with SNR below a certain minimum value, SNR$_{\mathrm{min}}$, from the test dataset (Fig.~\ref{fig:detector-snr-performance}). We observe a gradual improvement in performance with increasing SNR. For example, by considering only samples with $\text{SNR} > 4.0$, the false-positive rate is reduced from 35 to 6 occurences per hour, while maintaining a recall of 90\% and retaining more than 95\% of the upcalls in the test dataset. 

\section{\label{sec:4} Discussion}

The DNN classifier has been found to outperform the baseline LDA model, achieving recall and precision of $87.5\%$ and $90.2\%$ on the AB test dataset. %
Additionally, the DNN models trained on the combined datasets generally performed better than the models trained on the individual datasets, also when tested on the individual datasets. For example, models trained on AB consistently outperformed models trained exclusively on A, even when tested solely on A. In contrast, the baseline LDA model achieved worse performance when trained on combined datasets. This is an important observation, because it suggests that DNNs have the capacity to handle larger variance in the data, and indeed benefit from being trained on data with greater variance, producing models that are more robust to inter-dataset variability. %

On the other hand, we found that the DNN models generally performed poorly when trained on one dataset, but tested on another (e.g.\ trained on A, but tested on B). This behavior was not observed with the LDA models, whose less performant solution appear to be less sensitive to the training dataset. 

The quality and accuracy of the training datasets built as part of this work is limited by both the use of a classical time-frequency based detector to select candidate upcalls for expert validation and by human subjectivity in the validation step. Any bias in the selection or validation step will be reflected in the training dataset and hence affect the learning of the DNN. %
To explore the bias in the validation step, a second expert was tasked with reviewing all the 
incorrectly classified segments (false positives as well as false negatives) and an equally large number of correctly classified segments randomly sampled from the AB test dataset (cf.~Sec.~\ref{subsec:2:2}). %
The second expert flagged about half of the incorrectly classified segments as ``borderline'', implying that the expert considered these classifications as being highly uncertain. On the other hand, the second expert only flagged 9\% of the correctly classified segments as borderline. Removing the borderline cases from the test data improves the recall and precision by $2$\% and $5$\%, respectively. However, the second expert also changed some of the labels not considered to be borderline. Adopting the second expert's revised labels for the test data, the recall decreases by $6$\% while the precision increases by $2$\%. %
These changes in performance metrics testify to the difficulty of obtaining accurate annotations on PAM data. It would be interesting to investigate the inter-annotator variability in a more systematic and controlled manner than done here, but this is beyond the scope of the present study. (For example, it can be argued that the second expert may have been biased by prior knowledge of the labels proposed by the first expert and the DNN.) %

In order to obtain a realistic assessment of the performance that can be expected of the DNN model in a practical setting, we have tested the model's ability to identify upcalls in continuous acoustic data representative of the actual conditions required for a NARW upcall PAM DC system. In order to obtain an unbiased estimate of the detection performance, these data were subject to a complete manual validation not resorting to the use of a classical detection algorithm to select candidate upcalls. %
The best-performing model achieved a recall of 80\% while maintaining a precision above 90\%, corresponding to a false-positive rate of 5 occurences per hour for the chosen test dataset, while the ``average'' model only achieved a recall of 60\% for the same level of precision. However, by restricting our attention to upcalls with $\mathrm{SNR} \gtrsim 4.0$, the recall of the average model was increased to 85\% for the same level of precision while retaining over 95\% of the upcalls. %
Existing algorithms are capable of achieving similar levels of recall, but at the cost of a significantly higher false-positive rate \citep{simard2019, dclde2013}. %

Finally, we note that a related work entitled ``Deep neural networks for automated detection of marine mammal species''~\cite{shiu2020} has been published during the review of our paper. We would like to highlight the complementary nature of the two studies. While similar deployment techniques (surface buoys, bottom moorings) and acoustic recorders were used, \cite{shiu2020} considers acoustic data from several locations off the east coast of the US, whereas our work considers data from the Gulf of St.~Lawrence. Where \cite{shiu2020} provides a comparison of several network architectures, our work provides insights into the importance of dataset size and variance. Moreover, our work provides insights into recall variability with SNR. Although a direct comparison of the detection performances achieved in the two studies cannot be made since different test datasets were used, we note that the recall obtained by \cite{shiu2020} on continuous test data is somewhat higher than the recall obtained in our study (95\% vs.\ 87\% for a false-positive rate of 20~h$^{-1}$). However, the difference is within the range of variability that could be explained by differences in SNR in the test data.

\section{\label{sec:5} Conclusion}

In summary, we have demonstrated that DNNs can be trained to recognize NARW upcalls in acoustic recordings which have been made with different acoustic equipment and deployment types, and hence differ significantly in terms of signal amplitude and noise background. By training a DNN on a dataset comprised of about $4,000$ samples of NARW upcalls and an approximately equal number of negative samples, we achieved recall and precision of 90\% on a test dataset containing about 700 upcalls and a similar number of negatives. %
The DNN was observed to benefit from being trained on data with increased variance, suggesting that improved performance could be achieved by further expanding the variance of the training dataset. %
Using the DNN classifier, we implemented a simple detection algorithm, which exhibited good performance on continuous test data, achieving a recall of 80\% while maintaining a precision above 90\%. It would be interesting to explore still more sophisticated machine-learning approaches, most notably approaches that consider a wider temporal context, as done by the human experts, but this is beyond the scope of the present study and is left for future work. %
These results highlight the potential of DNNs for solving sound detection and classification tasks in underwater acoustics and motivate a community effort towards building larger and improved training datasets, especially for deployments with interfering noise events, which present more challenging acoustic conditions.

\begin{acknowledgments}

We would like to thank C.\ Hilliard for proofreading the manuscript. %
We are grateful to G.\ E.\ Davis, H.\ Johnson, C.\ T. Taggart, C.\ Evers, and J.\ Theriault for numerous discussions on NARW bioacoustics that supported and guided the development of the neural-network classifier. %
We thank Fisheries and Oceans Canada and the Natural Sciences and Engineering Research Council, Discovery Grant to YS, for the collection and preparation of the datasets. %
The NARW detection algorithm is a contribution of the Canadian Foundation for Innovation MERIDIAN cyberinfrastructure (\url{https://meridian.cs.dal.ca/}). %
This work was presented at the fifth International Meeting on The Effects of Noise on Aquatic Life held in Den Haag, July 2019.

\end{acknowledgments}

\appendix*
\section{Supplementary Material}

Upon manuscript acceptance we will provide the code necessary to reproduce the results presented in the paper including initialization and training of the neural network, prediction on test data, and computation of test metrics, along with all necessary training and test data. We will also provide more spectrogram samples (including expert annotation, model output, and SNR) and a complete diagram of the neural network architecture.

\end{document}